\documentclass[aps,prb,amsmath,amssymb,footinbib,showpacs,floatfix,twocolumn,superscriptaddress]{revtex4-2}
\usepackage{amsmath}
\usepackage{amssymb}
\usepackage{amsthm}
\usepackage{setspace}
\usepackage{graphicx}
\usepackage{braket}
\usepackage{mathrsfs}
\usepackage{times}
\usepackage{bm}
\usepackage[linktocpage=true,colorlinks=true,urlcolor=blue,citecolor=red]{hyperref}
\usepackage{amsfonts}
\usepackage{amsmath}
\usepackage{amssymb}
\usepackage{amsthm}
\usepackage{mathrsfs}
\usepackage{setspace}
\usepackage{xcolor}
\usepackage{braket}

\begin{document}

\newcommand{\muup}{\mu}

\title{Proposing one-sided tetrons as topological qubits for Majorana-based quantum computation}

\author{Tudor D. Stanescu}
\affiliation{Department of Physics and Astronomy, West Virginia University, Morgantown, WV 26506, USA}
\author{Sumanta Tewari}
\affiliation{Department of Physics and Astronomy,
Clemson University, Clemson, SC 29634}

\begin{abstract}
One-dimensional semiconductor-superconductor (SM-SC) hybrid nanowires with Rashba spin-orbit coupling and applied Zeeman field can support Majorana zero modes (MZMs) at the ends of the wire. In finite-length systems, residual wave-function overlap between the end modes produces an energy splitting. In a two-sided tetron qubit, composed of a pair of parallel Majorana wires coupled by a trivial SC backbone in the middle, this splitting limits the parity lifetime associated with Majorana pairs involving both wires and the coherence of the qubit, 
adversely affecting measurement-based braiding and readout operations. Here, we propose a \textit{one-sided} tetron made of four parallel Majorana wires connected at a common end by an SC backbone. In the topological phase, the free wire ends host four MZMs that define the topological qubit, while the Majoranas localized near the opposite ends hybridize through the backbone and acquire finite energy. 
To study the basic hybridization mechanism, we consider a simpler two-wire subsystem, i.e., half of a one-sided tetron, which we call a {\em doublon}. In a doublon, the backbone, or bridge, strongly hybridizes the two Majorana modes at the connected end and shifts them away from zero energy, while the modes at the free ends form a near-zero energy fermion.
We show that the splitting of the surviving MZM pair can fall by about two orders of magnitude compared to the splitting energies in a two-sided tetron. A low-energy model that includes a readout quantum dot coupled to the MZMs further shows that the bridge-induced hybridization energy suppresses charge-noise-induced transitions between low-energy eigenstates within a fixed total-parity sector, thereby increasing the parity lifetime. Thus, the one-sided tetron ensures a suppression of the Majorana overlap comparable to doubling the length of the corresponding two-sided tetron, without increasing the distance between MZMs or the size of the quantum dots used for implementing quantum operations. 
\end{abstract}

\maketitle

\section{Introduction}

\begin{figure*}[t]
\centering
\includegraphics[width=0.88\textwidth]{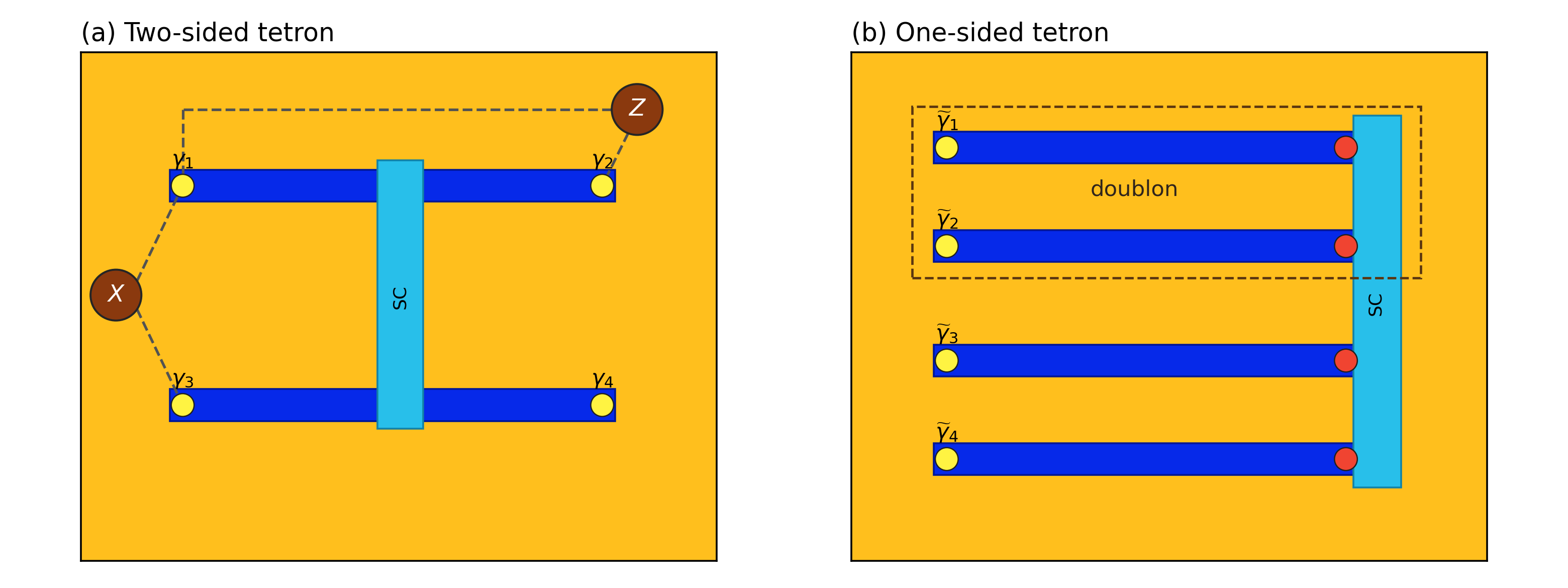}
\caption{Schematic representation of two different tetron geometries. 
The orange region represents a semiconductor heterostructure hosting a two-dimensional quantum well, blue corresponds to topological wires, and cyan represents the trivial SC backbone. The Majorana modes at the free ends are represented by yellow circles. (a) Two-sided tetron hosting the Majorana modes $\gamma_1,\gamma_2$ (upper wire) and $\gamma_3,\gamma_4$ (lower wire). The brown quantum dots coupled to the Majorana modes as indicated by the dashed lines realize distinct ($X$ and $Z$) readout loops.
(b) One-sided tetron with four arms terminating at one common backbone. Their free-end modes $\widetilde\gamma_1,\ldots,\widetilde\gamma_4$ encode a qubit. Red points indicate hybridized Majorana modes lifted to finite energy by the backbone. The boxed upper pair of arms is the doublon analyzed in this work (also see Fig. \ref{FIG1}).}
\label{FIGgeometry}
\end{figure*}

Majorana zero modes (MZMs), a class of exotic excitations in condensed matter systems with non-Abelian statistics, provide a nonlocal, topologically robust way to store and process quantum information and are a central ingredient of proposed topological quantum computation ~\cite{Kitaev_2001,Nayak2008,Moore1991,Nayak1996,Read2000,Kitaev2003}. Semiconductor-superconductor (SM-SC) heterostructures offer a practical route to this physics \cite{PhysRevB.77.220501,sau2010generic,oreg2010helical,lutchyn2010majorana,sau2010non,TEWARI2010219}. With spin-orbit coupling, induced superconducting pairing, and an externally applied Zeeman field, a SM-SC hybrid nanowire can enter a topological superconducting (TS) phase supporting MZMs at the wire-ends. This proposal has motivated extensive experimental activity in SM-SC heterostructures~\cite{ScienceMourik2012,Deng2012,NatPhysDas2012,Rokhinson_2012,churchill2013superconductor,finck2013anomalous,NatureAlbrecht2016,deng2016majorana,zhang2017ballistic,chen2017experimental,PRLNichele2017,albrecht2017transport,sherman2017normal,o2018hybridization,shen2018parity,vaitiekenas2018selective,Yu_2021,zhang2021,PhysRevB.107.245423, MicrosoftQuantumNatureReply2026, MicrosoftAzureQuantumNature2025,TetronLifetimes,MicrosoftQuantumPb2026}. In recent experiments, researchers have developed interferometric parity readout on so-called tetron qubit devices \cite{Tetron_2017}, which are based on this platform \cite{MicrosoftAzureQuantumNature2025, TetronLifetimes, MicrosoftQuantumPb2026}. The tetron devices consist of two parallel SM-SC nanowires, connected in the center by a trivial superconducting bridge or backbone (Fig.~\ref{FIGgeometry}). Because the bridge or the backbone is at the center, this type of tetron, henceforth called "two-sided tetron" [Fig.~\ref{FIGgeometry}(a)], has two free wire ends on each side.  When the SM-SC nanowires are tuned into the TS phase using Zeeman fields and voltage gates, four MZMs emerge, two on each side of the so-called sideways "H" structure, forming a topological qubit \cite{ MicrosoftAzureQuantumNature2025, TetronLifetimes, MicrosoftQuantumPb2026, Tetron_2017}.
Specifically, the four end MZMs are $\gamma_1,\gamma_2$ on the upper wire and $\gamma_3,\gamma_4$ on the lower wire [see Fig. \ref{FIGgeometry}(a)].
The parity of the fermion corresponding to the Majorana pair $(\gamma_j,\gamma_k)$, which is defined by $c^{\dagger}_{jk}=\frac{1}{2}(\gamma_j-i\gamma_k)$ and $c_{jk}=\frac{1}{2}(\gamma_j+i\gamma_k)$, is given by the eigenvalues of the operator $i\gamma_k\gamma_j = 1-2n_{jk}$, where $n_{jk}=c_{jk}^{\dagger}c_{jk}$ is the fermion occupation number. Specifically, even and odd parities correspond to eigenvalues $+1$ $(n_{jk}=0)$ and $-1$ $(n_{jk}=1)$, respectively. 
Within a fixed total-parity sector of the two-sided tetron, we choose
\begin{equation}
 X=i\gamma_1\gamma_3,\qquad Z=i\gamma_1\gamma_2.
 \label{paulis}
\end{equation}
The $X$ parity refers to the occupation number of the MZMs at the two left ends, whereas the $Z$ parity belongs to the MZMs at the two ends of the upper wire. Since the operators share $\gamma_1$ while $\gamma_2$ and $\gamma_3$ anticommute, the $X$ and $Z$ operators anticommute: a state with a definite $Z$ parity cannot have a definite $X$ parity. The lower-wire parity $i\gamma_3\gamma_4$ is determined by $Z$ and the fixed total parity $M$ given by the product $M=(i\gamma_1\gamma_2)(i\gamma_3\gamma_4)=\pm 1$. The trivial backbone is designed to make direct single-electron tunneling between the wires very small.

The $Z$ and $X$ parities can be measured with interferometers that include quantum dots coupled to the tetron \cite{MicrosoftAzureQuantumNature2025,TetronLifetimes, MicrosoftQuantumPb2026}. To measure $X$, a dot is coupled to the left Majoranas, $\gamma_1$ and $\gamma_3$; to measure $Z$, a different dot is coupled to the two ends of the upper wire, i.e., to $\gamma_1$ and $\gamma_2$. In each case, the quantum dot and a superconducting path, e.g., the left halves of the wires plus the SC backbone for the $X$ parity measurement or the top wire for the $Z$ measurement [see Fig.~\ref{FIGgeometry}(a)], form a closed loop that is threaded by a magnetic flux. Interference between the two paths forming the closed loop produces a parity-dependent shift of the dot energy. The quantum capacitance of the dot is the curvature of this energy with respect to the dot gate voltage, conventionally $C_Q=-\partial^2 E_g/\partial V_g^2$ after accounting for the gate lever arm. Observation of a fermion parity-dependent quantum capacitance, periodic in the threaded flux with period $\frac{h}{e}$, is a necessary condition for the existence of MZMs at the ends of the tetron system. 
In recent experiments, researchers have been able to resolve this parity-dependent periodic capacitance response of the quantum dot \cite{MicrosoftAzureQuantumNature2025, TetronLifetimes, MicrosoftQuantumPb2026,  VigneauAPR2023}. 
Repeated capacitance measurements of the quantum dots yield a record of $X$ or $Z$ outcomes, each with a bimodal distribution of two capacitance values corresponding to opposite parities. The switching times between the two capacitance values associated with opposite parities, which are extracted from the measurement records, are called parity lifetimes and are denoted $\tau_X$ and $\tau_Z$, respectively.
Since for measurement-based braiding and reliable gate operations the parity associated with a Majorana pair must remain unchanged following the corresponding parity measurement, through the switching between different measurement settings, long-enough parity lifetimes are required for the success of Majorana-based quantum computation.

Parity lifetimes, e.g., $\tau_Z$ and $\tau_X$, can differ because they are affected differently by various noise sources and by the presence of a finite Majorana overlap. 
In each finite-length wire, the end-mode Majorana wave functions necessarily overlap. The leading contribution of this hybridization to the two-sided tetron Hamiltonian is given by,
\begin{equation}
 H_{\rm ov}=iE_{12}\gamma_1\gamma_2+iE_{34}\gamma_3\gamma_4.
 \label{overlap_intro}
\end{equation}
Both terms in $H_{\rm ov}$ commute with $Z=i\gamma_1\gamma_2$ but anticommute with $X=i\gamma_1\gamma_3$. For a fixed total fermion parity $M=\pm 1$, $H_{\rm ov}=(E_{12}+M~\! E_{34})Z$.
Therefore, for an isolated qubit (i.e., without a quantum dot connected to it), a freshly prepared $X$ eigenstate undergoes coherent precession on a scale set by $\hbar/|E_{12}\pm E_{34}|$. Although this is by itself coherent evolution, the corresponding timescale can be very short, e.g., $\sim 3~$ns for roughly one radian of phase accumulation for energy splittings $E_{12}\sim E_{34} \sim 0.1$ $\muup$eV.
On the other hand, the stochastic parity switching quantified by $\tau_X$, which is exacted from measured capacitance time traces, involves not only the tetron, but also the measurement dot and its environment. Consequently, $\tau_X$ depends both on the Majorana overlap and on the noise strength (e.g., the potential fluctuations) affecting the measuring dot, and, typically, corresponds to a different timescale. 
For example, if the Majorana-quantum dot coupling is much larger than the Majorana splitting energy, the stochastic parity switching times can be of the order of tens of microseconds for $E_{12}\sim E_{34} \sim 0.1$ $\muup$eV.
By contrast, the noise associated with measuring the $Z$ parity mainly involves quasiparticles tunneling from the external environment,  with single-electron transfer from the other wire through the trivial backbone being negligible \cite{MicrosoftAzureQuantumNature2025, TetronLifetimes, MicrosoftQuantumPb2026}. Thus, $\tau_Z$ is practically independent of the Majorana hybridization, e.g., the splitting energy $E_{12}$.

The contrast between the two types of parity measurements has been observed experimentally. Repeated measurements on an InAs--Al two-sided tetron gave $\tau_X=14.5\pm0.3~\mu\mathrm{s}$ and $\tau_Z=12.4\pm0.4~\mathrm{ms}$; the reported interpretation associates the shorter $\tau_X$ parity switching time scale with the presence of a significant intrawire overlap, while the much longer (by about three orders of magnitude) $\tau_Z$ timescale is controlled by external quasiparticle poisoning \cite{TetronLifetimes}. A more recent InAs--Pb tetron showed a characteristic $Z$ parity-switching time near $20~\mathrm{s}$, with some intervals reaching a minute \cite{MicrosoftQuantumPb2026}. However, the  $X$-loop lifetime only increased to approximately $180~\mu\mathrm{s}$ (unpublished). Relative to the Al-device, that would represent an enhancement by more than three orders of magnitude in the quasiparticle poisoning-sensitive parity lifetime, $\tau_Z$, but  only about one order of magnitude in $\tau_X$. 

The likely cause of this huge disparity between $\tau_X$ and $\tau_Z$, even after the introduction of Pb as the parent SC, is a substantial Majorana overlap within the two nanowires that constitute the two-sided tetron device. In turn, the likely source of this overlap is the presence of disorder sufficiently strong to push the control parameters consistent with the presence of MZMs into a regime characterized by length scales comparable to the size of the device.
While this may be consistent with the possibility of operationally defining a topological phase and corresponding Majorana modes, it may not be enough to ensure any of the practical potential advantages associated with topological protection. To overcome this critical problem, as well as the issue associated with the fast coherent precession of an $X$ parity state of a tetron decoupled from QDs, one can try to suppress the Majorana hybridization by reducing the disorder strength (e.g., by optimizing the materials growth and device fabrication) or by improving the relevant effective parameters, such as spin-orbit coupling, g-factor, superconducting gap, etc., (e.g., by exploring new materials). However, these are slow, tedious processes that may take a long time to generate positive outcomes. Another possible approach is to use longer nanowires, to enhance the separation of the Majorana modes and reduce their hybridization. This, however, is likely to generate new technical issues. For example, the topological gap protocol \cite{PhysRevB.107.245423} currently used to tune the wires into the topological regime may not work for long wires, if the nonlocal conductance signal becomes too weak. More importantly, a Pauli-$Z$ measurement, which requires coupling to a quantum dot as long as the device itself [see Fig. \ref{FIGgeometry}(a)], may be difficult if the gaps separating the quantum dot energy levels become too small as a result of a large quantum dot size. 

In this paper, we explore an alternative solution based on the \emph{one-sided tetron} design, which can be implemented using
the current state-of-the-art. Our proposed one-sided tetron has four parallel topological arms joined at their right ends by a trivial SC backbone [Fig.~\ref{FIGgeometry}(b)]. Its four free left ends host the MZMs $\widetilde\gamma_1,\ldots,\widetilde\gamma_4$ that encode a qubit. The pair parities, $\widetilde X=i\widetilde\gamma_1\widetilde\gamma_2$ and $\widetilde Z=i\widetilde\gamma_1\widetilde\gamma_3$, for example, which anticommute, can be measured after realizing appropriate connections with quantum dots located near the left end of the device. Here, we do not model a complete four-arm measurement circuit. Instead, we focus on the upper two arms -- which we dub as a \emph{doublon} [see Fig. \ref{FIGgeometry}(a) and the details in Fig.~\ref{FIG1}(a)] -- and investigate the hybridization of the Majorana modes in this device.  
More specifically,  we address the following key questions: (i) How efficiently can one couple the right Majorana pair (M2-M4) through a superconducting bridge? (ii) What is the fate of the left Majorana pair (M1-M3) after M2-M4 hybridize and
acquire a finite energy $\delta$? (iii) What is the parity lifetime associated with the left Majorana pair (M1-M3) in the presence of a finite hybridization energy $\delta$? To address the first two questions, we use a two-dimensional effective tight-binding model for both a clean device and a system with strong disorder. We show that, upon coupling the Majorana pair M2-M4 through the bridge, the resulting bound state acquires an energy $\delta$ having values on the order of $10^1$ $\mu$eV over a significant range of parameters. At the same time, the splitting energy of
the “surviving” left Majorana pair, M1-M3, collapses toward
zero. As a result, a doublon of length $L_x$ can be seen as a
“folded” wire of length $2L_x$ having a partially-separated Andreev bound state or the so-called quasi-Majorana \cite{moore2018two, Moore2018, SciPostVuik2019, stanescu2019robust} of energy $\delta$ in the middle. We note that the
doublon model can be converted into a two-sided tetron
model by simply moving the bridge to the middle of the wires
and making it very narrow. The narrowness of the bridge and its placement in the middle of the wire, with the MZM wave functions peaked at the ends of the wire, result in a negligible inter-wire Majorana coupling, in particular an M2-M4 hybridization
energy $\delta \sim 0$.
For the lifetime calculations, we use a simplified low-energy effective model \cite{TetronLifetimes}. We find that the parity lifetime associated with the left Majorana pair (M1-M3) increases
dramatically with the hybridization energy $\delta$ of the right pair
(M2-M4) for the value of $\delta$ above a certain threshold (of order $10^1$ $\mu$eV). In our calculations, the parity lifetime of the left Majorana pair  reaches values that are two or three orders of
magnitude larger than the lifetime for $\delta = 0$, i.e., the parity
lifetime characterizing the Pauli-$X$ measurement on the corresponding two-sided tetron.

Finally, we reiterate that
the only difference between a doublon and a two-sided tetron
is the position and, possibly, the size of the superconducting bridge. In a doublon, the role of the bridge is to couple
the Majorana modes M2 and M4, therefore it is placed at the
right end of the structure and should have geometric parameters ($W_x, W_y$ ) that maximize this coupling. On the other
hand, in the two-sided tetron, the bridge should not generate
any significant coupling between the wires, particularly between the top (M1-M2) and bottom (M3-M4) Majorana pairs,
therefore it is placed in the middle of the structure, where the
Majorana modes have small amplitudes.

The remainder of this paper is organized as follows. In Section II we describe the proposed device and the theoretical
models used to characterize its properties. The results of the
numerical calculations are presented in Section III, starting
with the clean case (Sec. III A) and continuing with the disordered device (Sec. III B). The parity lifetime calculations are
discussed in Sec. III C. We present our discussion and conclusion in Sec. IV. 



\section{Device and theoretical model} \label{SecII}  

\begin{figure}[t]
\begin{center}
\includegraphics[width=0.48\textwidth]{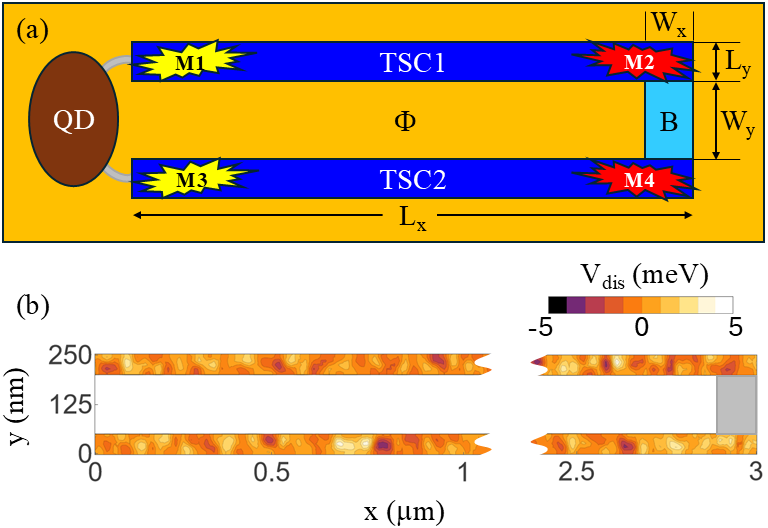}
\end{center}
\vspace{-2mm}
\caption{(a) Schematic representation of a doublon coupled to a quantum dot (QD). The doublon consists of two hybrid nanowires of length $L_x$ defined by thin superconducting films (blue) proximity coupled to a 2D electron gas hosted by a semiconductor structure (orange). The right ends of the wires are coupled through a superconducting bridge (B), while a magnetic flux $\Phi$ is threaded through the loop defined by the doublon and the quantum dot. The unproximitized semiconductor regions are depleted using top gates (not shown). In the topological superconducting phase the device hosts Majorana modes (M1-M4). (b) Spatial profile of the disorder potential within the wire regions. The amplitude of the disorder potential is $V_0 =\sqrt{\langle V_{\rm dis}^2\rangle} = 1.5~$meV, while the average value is $\langle V_{\rm dis}\rangle = 0$.}
\label{FIG1}
\vspace{0mm}
\end{figure}

The proposed doublon is shown in Fig. \ref{FIG1}(a). Two hybrid wires are formed by proximity coupling a semiconductor two-dimensional electron gas to thin superconducting films. A superconducting bridge joins their right ends. Gates deplete the semiconductor outside the intended wire and bridge regions. A magnetic field along the wires drives each wire toward the topological regime. Before the wires are joined, the top wire hosts M1 and M2 and the bottom wire hosts M3 and M4. The bridge allows M2 and M4 to overlap and form a finite-energy Andreev state \cite{moore2018two, Moore2018, SciPostVuik2019, stanescu2019robust}. M1 and M3 remain at the two open ends. Their joint parity can be read by coupling both ends to a quantum dot and measuring its flux-dependent quantum capacitance \cite{Tetron_2017,PRBPlugge2017,Munk2020,MicrosoftAzureQuantumNature2025, MicrosoftQuantumPb2026, TetronLifetimes,  PRBSau2025, StanescuTewariPRB2026}. In the doublon the bridge sits at the wire ends, where the M2 and M4 amplitudes are large. In a two-sided tetron the backbone sits near the middle, where these amplitudes are small.

To characterize the low-energy properties of the doublon, we use an effective tight-binding model for the semiconductor 2DEG, with parameters renormalized by the proximity effect \cite{Paudel2025}. More specifically, after integrating out the superconductor degrees of freedom and using the static approximation for the resulting self-energy, a spin and particle-hole block of the effective
Hamiltonian corresponding to lattice indices $i(j) \in A$ and $j(i)\in B$ takes the form
\cite{Paudel2025}
\begin{eqnarray}
[H_{eff}]_{ij} &=&  \sqrt{Z_A Z_B}~[\widetilde{H}_{SM}]_{ij}-\Delta_A ~\!\sigma_y\tau_y ~\!\delta_{ij},   \label{Heff}
\end{eqnarray}
where $i$ and $j$ label sites on a 2D square lattice with lattice constant $a$ and $A,B\in \{{\rm SM, SC, B}\}$ designate different regions of the device corresponding to the unproximitized semiconductor (SM), the superconducting wire regions (SC), and the superconducting bridge region (B). In Eq. (\ref{Heff}), $\sigma_y$ and $\tau_y$ are Pauli matrices associated with the spin and particle-hole degrees of freedom, respectively; $\Delta_A$ is the induced superconducting gap in region $A$, i.e., $\Delta_{SM} =0$, $\Delta_{SC} = \gamma~\!\Delta_0/(\gamma +\Delta_0)$, and $\Delta_{B} = \gamma_B~\! \Delta_0/(\gamma_B +\Delta_0)$, with $\Delta_0$ being the parent SC gap and $\gamma$ ($\gamma_B$) representing the effective SM-SC coupling within the wire (bridge) region;  $Z_A$ is the quasiparticle weight inside the semiconductor in region A, i.e., $Z_{SM}=1$, $Z_{SC}=\Delta_0/(\gamma+\Delta_0)$, and $Z_{B}=\Delta_0/(\gamma_B+\Delta_0)$. Finally, $\widetilde{H}_{SM}$ is the semiconductor Hamiltonian matrix in the Bogoliubov-de Gennes form, $\widetilde{H}_{SM}= H_{SM}^{}(1+\tau_z)/2 -H_{SM}^*(1-\tau_z)/2$, with $H_{SM}$ corresponding to the bare second quantized SM Hamiltonian 
\begin{eqnarray}
 H_{SM} &=& \sum_{i}\left(4t-\mu +V_i\right) \hat{c}_{i}^\dagger \hat{c}_{i}^{} -t\!\sum_{i,\delta}{\hat c^\dag_i \hat c_{i+\delta}^{}} + \Gamma \!\sum_{i} \hat{c}_{i}^\dagger \sigma_x \hat{c}_{i}^{} \nonumber \\
 &+& \frac{i \alpha_R}{2}\sum_{i}[\hat c^\dag_{i +\delta_x} \hat \sigma_y \hat c_i^{} - \hat c^\dag_{i +\delta_y} \hat \sigma_x \hat c_i^{} + \text h.c.], \label{HSM}
\end{eqnarray}
where  $i=(i_x,i_y)$ labels sites on a square lattice with lattice constant $a$,  $\delta=(\delta_x,\delta_y)$ are nearest neighbor vectors, and $\hat c_i^\dag=(\hat c^\dag_{i\uparrow}, \hat c^\dag_{i\downarrow})$,  $\hat c_i=(\hat c_{i\uparrow}^{}, \hat c_{i\downarrow}^{})^T$ are the electron creation and annihilation operators, respectively. The parameters of the tight-binding Hamiltonian are: the nearest-neighbor hopping amplitude, $t = {\hbar^2}/{2m^* a^2}$, with $m^*$  being the effective mass,  the chemical potential, $\mu$, the Rashba spin-orbit coupling strength, $\alpha_R$, and the (half) Zeeman splitting, $\Gamma=\frac{1}{2} g \mu_B B_x$, generated by an external magnetic field applied along the $x$-direction (i.e., parallel to the wires). The position-dependent potential has two components, $V_i=V_{\rm conf}(i) + V_{\rm dis}(i)$, with $V_{\rm conf}$ representing the potential that confines the electrons within the proximitized regions and $V_{\rm dis}$ being a random disorder potential. For simplicity, in the calculations we assume that $V_{\rm conf}$ is zero under the superconductors (i.e., in regions SC and B) and is infinitely large in the SM region, except two narrow strips ($20~$nm wide) on both sides of the bridge, where $V_{\rm conf}=0.5~$eV. We use this softer confinement to better characterize the coupling of the right Majoranas through the bridge, but we emphasize that, for an accurate quantitative description and for optimizing the structure, a more detailed modeling may be necessary. The disorder potential used in the calculations, which is shown in Fig. \ref{FIG1}(b), is characterized by zero average, $\langle V_{\rm dis}\rangle = 0$, and by an amplitude $V_0=1.5~$meV, where $V_0^2 =\langle V_{\rm dis}^2\rangle$, with $\langle\dots\rangle$ designating the position average.

The values of the model parameters used in the numerical calculations are as follows. 
The lattice constant of the square lattice is $a=10~$nm; the hopping parameter is $t=12.7~$meV, which corresponds to an effective mass $m^*=0.03m_0$, with $m_0$ being the free electron mass; the Rashba spin-orbit coefficient is $\alpha_R =2~$meV, corresponding to $\alpha_R\cdot a =200~$meV$\cdot$\AA; the parent SC gap is $\Delta_0=0.3~$meV; the effective SM-SC coupling within the wire regions (SC) is $\gamma=0.35~$meV, which corresponds to an intermediate coupling regime, $\gamma \sim\Delta_0$, while for the bridge region (B) we have $\gamma_B=1.4~$meV  (i.e., strong coupling); the width of the SC wires is $L_y=50~$nm and we consider different values of the wire length ($L_x$), bridge length ($W_y$), and bridge width ($W_x$), which are specified below; the chemical potential, $\mu$, and the Zeeman field, $\Gamma$, are used as control parameters. We note that the effective SM-SC coupling parameters $\gamma$ and $\gamma_B$ correspond to values of the quasiparticle weight and induced gap in the wire and bridge regions $Z_{SC}=0.46$, $\Delta_{SC} = 0.16~$meV and $Z_B=0.18$, $\Delta_B=0.25$, respectively. The strong SM-SC coupling characterizing the bridge region was chosen to ensure that the bridge remains in a trivial superconducting state for all Zeeman field values used in the calculation ($\Gamma \lesssim 1.3~$meV). 

We note that the suppression of the free-end (M1--M3) splitting induced by the M2--M4 hybridization can be obtained without using a detailed semiconductor model. Indeed, consider the Majorana Hamiltonian \begin{equation}
H_{\rm M}= H_{\rm ov}+ H_{\delta}=iE_{12}\gamma_1\gamma_2+iE_{34}\gamma_3\gamma_4+i\delta\gamma_2\gamma_4,
\label{Eq:H_doublon}
\end{equation} 
where $\delta$ is the bridge-induced coupling. When $|\delta|$ is much larger than $|E_{12}|$ and $|E_{34}|$, the joined-end pair is costly to excite. Eliminating it perturbatively leaves a coupling between the open-end modes of magnitude $|E_{\rm free}|\sim |E_{12}E_{34}/\delta|$, up to convention-dependent factors. A coupling that was first order in the overlap of a single wire is thus replaced by a second-order path through both wires. The two-dimensional calculation below tests this picture without assuming that the three couplings are independent parameters.

The parity lifetime calculations were done using a low-energy effective model \cite{TetronLifetimes} that includes four Majorana modes (M1--M4, described by $\hat{\gamma}_1$--$\hat{\gamma}_4$) and one quantum dot state described by the creation and annihilation operators $\hat{d}^\dagger$, $\hat{d}$ (see Fig. \ref{FIG1}).  Defining the annihilation operators for the left and right Dirac fermions as $\hat{c}_L^{}=\frac{1}{2}(\gamma_1 + i\gamma_3)$ and $\hat{c}_R^{}=\frac{1}{2}(\gamma_2 + i\gamma_4)$, respectively, the basis for the effective Hamiltonian is given by the states $|n_D~n_L~n_R\rangle$, where $n_\alpha$ is the occupation number of the fermion mode $\alpha$, corresponding to a given total parity $M$. Explicitly, the basis of the even parity sector ($M=-1$) is $\{|101\rangle, |011\rangle, |110\rangle, |000\rangle \}$, while for $M=+1$ we have $\{|100\rangle, |010\rangle, |111\rangle, |001\rangle \}$. The corresponding effective Hamiltonian, in the presence of a magnetic flux $\Phi$ through the closed loop formed by the QD, two wire arms and the SC bridge, takes the form
\begingroup
\setlength{\arraycolsep}{2pt}
\begin{equation}
\cal{H}=
\left(\begin{array}{cccc}
\frac{E_D}{2}\!-\!M\delta & t_1\!-\!it_3 e^{-i\Phi} &-i\epsilon_M^- & 0 \\
t_1\!+\!it_3 e^{i\Phi} & -\frac{E_D}{2}\!-\!M\delta & 0 & i\epsilon_M^+\\
i\epsilon_M^- & 0 & \frac{E_D}{2}\!+\!M\delta & t_1\!+\!it_3 e^{-i\Phi} \\
0 & -i\epsilon_M^+ & t_1\!-\!it_3 e^{i\Phi} & -\frac{E_D}{2}\!+\!M\delta
\end{array}\right), \label{HM}
\end{equation}
\endgroup
where $\epsilon_M^\pm = E_{12} \pm E_{34} M$, with $E_{12}$ and $E_{34}$ representing the intra-wire Majorana energy splittings, $\delta\equiv E_{24}$ is the hybridization energy of the right Majoranas (through the bridge), $E_D$ is the energy of the dot state, $t_1$ and $t_3$ are the couplings between the left Majorana modes and the quantum dot, and $\Phi$ is the magnetic flux threaded through the doublon-QD loop in units of $\hbar/e$. For a given total parity M, the zero temperature parity lifetime $\tau_M$ associated with the left Majorana pair can be estimated based on the transition rate between the lowest energy states of ${\cal H}$, $|\chi_-^M\rangle$ and $|\chi_+^M\rangle$, which is approximately \cite{Munk2020,TetronLifetimes}   
\begin{equation}
\tau_M^{-1} = S_{n}(0)~\! |\langle \chi_+^M|H_{noise}| \chi_-^M\rangle|^2, \label{tauM}
\end{equation}
where $H_{noise}={\rm diag}(+1/2, -1/2, +1/2, -1/2)$ is the dimensionless noise Hamiltonian describing charge noise on the quantum dot with a stochastic amplitude characterized by the spectral function $S_n(\omega)=S_n(0)/(1+e^{-\omega/k_BT})$ and, at zero temperature, $S_n(0)$ is the effective noise strength. We also define the average lifetime as $\tau = 2/(\tau_+^{-1} + \tau_-^{-1})$. 
The parameter values used in the numerical calculations are $E_{12}=0.15~\muup$eV, $E_{34}=0.85~\muup$eV, $t_1=2~\muup$eV,  $t_3=3~\muup$eV, and $S_n(0) = 0.05~$GHz. The other parameter values are provided in Sec. \ref{SecIIIC}.

We emphasize that Eq.~(\ref{tauM}) estimates the switching time between the low-energy states within one conserved total-parity sector. It describes the component of the parity lifetime caused by charge noise on the quantum dot; it does not include the (much slower) parity switch processes induced by quasiparticle poisoning, when external quasiparticles change the total parity of the system. In particular, the quoted $\tau_M$ should not be equated with the free-precession time of an isolated tetron or with the experimental $\tau_Z$, but is closely related (in terms of dominant parity switching mechanism) to the $\tau_X$ time characterizing a two-sided tetron. 

\section{Results} \label{SecIII}  

By solving numerically the effective model defined by Eqs. (\ref{Heff}) and (\ref{HSM}), we first demonstrate that in a clean doublon device the presence of the bridge results in the right Majorana pair (M2-M4) hybridizing and acquiring a finite energy on the order of $0.03\Delta$--$0.3\Delta$, while the energy splitting of the left Majorana pair (M1-M3) gets substantially reduced (Sec. \ref{SecIIIA}). In Sec. \ref{SecIIIB} we confirm that this physics survives in the presence of disorder. The impact of the M2-M4 hybridization on the parity lifetime associated with the M1-M3 Majorana pair is evaluated in Sec. \ref{SecIIIC} based on the simplified model defined by Eqs. (\ref{HM}) and (\ref{tauM}).

\subsection{Clean device} \label{SecIIIA}  

\begin{figure}[t]
\begin{center}
\includegraphics[width=0.48\textwidth]{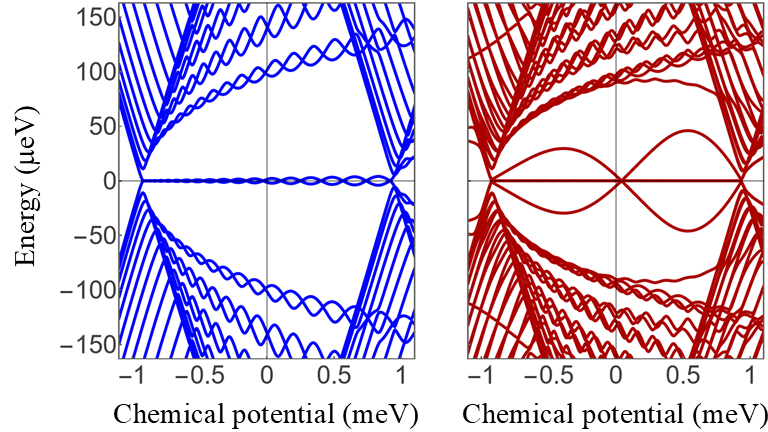}
\end{center}
\vspace{-2mm}
\caption{Comparison between the energy spectra as functions of the chemical potential  for a clean wire (left) and a doublon obtained by coupling two such wires (right). The Zeeman field is $\Gamma=1~$meV and the geometric parameters of the device are $L_x= 1.5~\muup$m, $L_y=50~$nm, $W_x=80~$nm, and $W_y=150~$nm. For the decoupled wires (left), the partial overlap of the Majorana modes (e.g., M1 and M2 or M3 and M4) results in energy splitting oscillations with an amplitude of up to $4~\muup$eV. By coupling the right ends of the wires (right panel), the right Majorana modes (M2 and M4) hybridize and acquire a finite energy of up to $45~\muup$eV, while the energy splitting of the ``surviving'' left Majorana modes (M1 and M3) is reduced by about two orders of magnitude to less than $0.04~\muup$eV. For comparison, a clean wire with $L_x=3~\muup$m (all other parameters being the same) is characterized by Majorana energy splitting with an amplitude of up to $0.09~\muup$eV.}
\label{FIG2}
\vspace{0mm}
\end{figure}

In Fig. \ref{FIG2}, we compare the energy spectrum (as a function of the chemical potential) of a clean, relatively short ($L_x=1.5~\muup$m) wire (left panel) with the energy spectrum of the doublon obtained by coupling two such wires (right panel). Note the (near) closing of the bulk gap at the topological quantum phase transitions corresponding to $\mu = \pm\sqrt{\Gamma^2-\gamma^2}\approx \pm 0.94~$meV and the significant energy splitting oscillations (with amplitude up to $4~\muup$eV) associated with the partial overlap of the Majorana modes within the wire (left). The spectrum of the corresponding doublon (right) is characterized by a reduction of the energy splitting of the ``surviving'' left Majorana pair by about two orders of magnitude, while the right Majoranas hybridize and acquire a finite energy having values larger than $10~\muup$eV over most of the chemical potential range consistent with the topological phase (see the second-lowest energy mode). Comparing the left and right panels also indicates that the bulk spectrum of the wires is weakly affected by the presence of the bridge, but the doublon supports additional low-energy modes localized near the bridge (e.g., the third lowest energy mode near $\mu=0.5~$meV). Finally, we point out that the splitting energy of the right Majorana pair is practically independent of the device length for $L_x\gtrsim 1.5~\muup$m, since M2 and M4 are localized modes with amplitudes near the right end of the wires that depend very weakly on $L_x$. 

\begin{figure}[t]
\begin{center}
\includegraphics[width=0.48\textwidth]{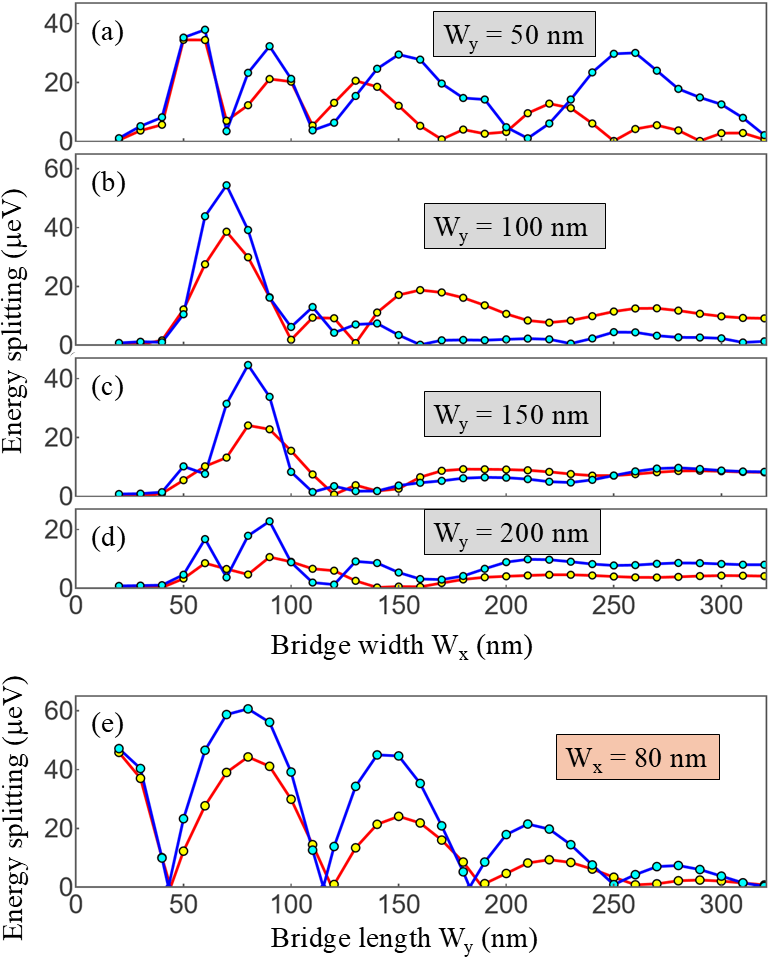}
\end{center}
\vspace{-2mm}
\caption{(a-d) Dependence of the energy characterizing the (partially separated) Andreev bound state obtained by hybridizing the right Majorana modes, M2--M4, on the width of the bridge ($W_x$) for different values of the bridge length ($W_y$). The length of the device is $L_x=2~\muup$m, while the control parameters are $\mu=0.3~$meV, $\Gamma=0.7~$meV (yellow circles and red lines) or $\mu=0.6~$meV, $\Gamma=1~$meV (cyan circles and blue lines). (e) Dependence of the energy characterizing the hybrid mode M2--M4 on the length $W_y$ of the superconducting bridge for a doublon with $W_x=80~$nm, all other parameters being the same as in (a-d).}
\label{FIG3}
\vspace{0mm}
\end{figure}

A natural question concerns the dependence of the splitting energy $\delta$ of the right Majorana pair on the parameters of the bridge. Here, we focus on the geometric parameters, i.e., the bridge length, $W_y$, and width, $W_x$, and show that the large energy splitting condition $\delta\gtrsim 10~\muup$eV can be realized within a reasonably large parameter range. We emphasize that an optimization study aiming at maximizing $\delta$ should probably involve a more detailed device model. In Fig. \ref{FIG3} we show the energy splitting as a function of $W_x$ for different bridge lengths [panels (a-d)] and as a function of $W_y$ for a fixed width $W_x=80~$nm [panel (e)]. First, we notice the non-monotonic dependence of $\delta$ on the bridge parameters. Second, we point out the clear trend in (e): the energy splitting has an oscillatory dependence on the bridge length, with an amplitude that decreases with $W_y$. Finally, we note that panels (a)--(d) consistently show large values of the splitting energy for bridge widths within the range $60-100~$nm. This length scale, which corresponds to $2\pi/k_F$, where $k_F$ is the characteristic Fermi wave vector of the system for the values of the control parameters used in the calculation, provides a useful criterion for identifying the values of $W_x$ consistent with large energy splittings for any given regime of control parameters, e.g., the regime likely to support a topological phase.    

The proof-of-concept calculation presented above shows that, by coupling two topological superconducting wires in a doublon device, one obtains a pair of robust Majorana zero modes localized near one end of the device (e.g., the left end) and a partially separated Andreev bound state consisting of a pair of hybridized Majoranas localized near the opposite end and having finite energy $\delta$. The splitting  energy $\delta$ represents a significant fraction of the topological gap (e.g., $\delta \gtrsim 10~\muup$eV) within a significant range of device and control parameters.

\subsection{Disordered device} \label{SecIIIB}  

Our next task is to demonstrate that the physics described above survives in the presence of (relatively) strong disorder. To this end, we consider two wires of length $L_x=3~\muup$m in the presence of a disorder potential with a spatial profile as shown in Fig. \ref{FIG1}(b) and strength $V_0=1.5~$meV. In general, the presence of such disorder potential is expected to result in a fragmentation of the topological phase and a shift in parameter space toward larger values of the chemical potential and Zeeman field \cite{RoyJaiswalStanescuTewariPRB2024,Eissele2025}. The specific topological areas in the $\Gamma-\mu$ plane associated with the two wires are, in general, different, since they correspond to different disorder realizations, but their typical location within the control parameter space is similar \cite{RoyJaiswalStanescuTewariPRB2024,Eissele2025}. Consequently, if the disorder strength is consistent with the presence of relatively large topological regions (within the relevant Zeeman field range), one can identify control parameter regions that support (operationally defined) topological superconductivity in both wires, possibly after tuning the chemical potential in one of the wires \cite{PhysRevB.107.245423}.  

\begin{figure}[t]
\begin{center}
\includegraphics[width=0.48\textwidth]{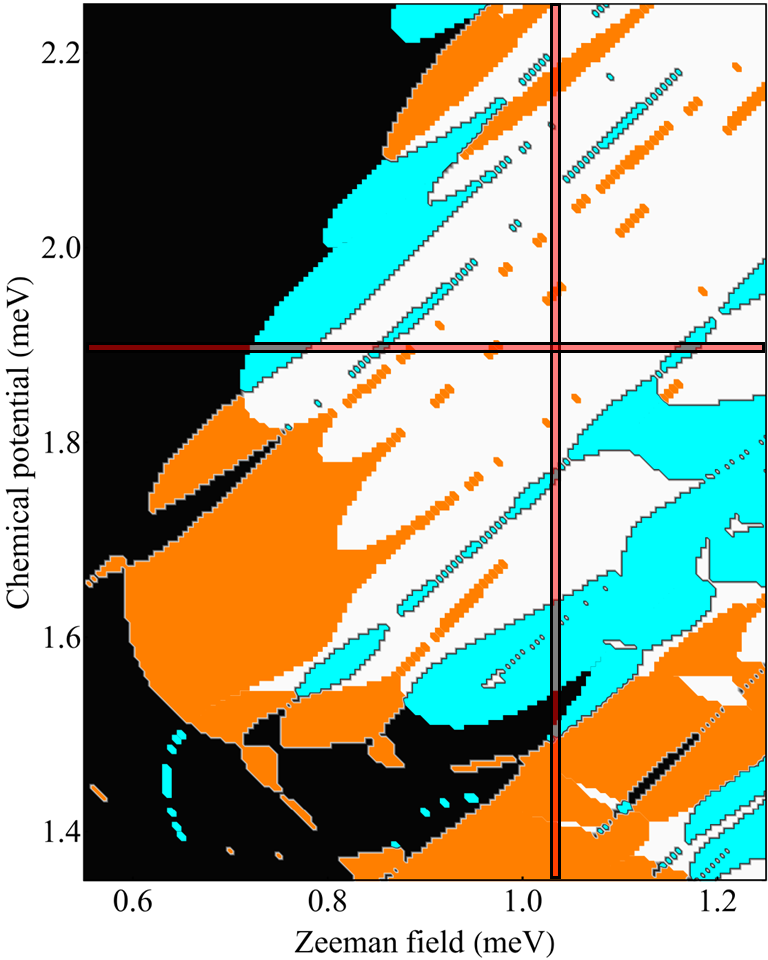}
\end{center}
\vspace{-2mm}
\caption{Topological phase diagrams characterizing two uncoupled wires of length $L_x=3~\muup$m in the presence of the disorder potential shown in Fig. \ref{FIG1}(b). The black region corresponds to both wires being in a trivial superconducting phase, while within the white region both wires are topologically nontrivial. The cyan (orange) areas indicate that the bottom (top) wire is topological, while the top (bottom) wire is in a trivial superconducting state. The energy spectra along the vertical and horizontal cuts marked by red lines are provided in Fig. \ref{FIG5} and Fig. \ref{FIG6}, respectively.}
\label{FIG4} 
\vspace{0mm}
\end{figure}

Figure~\ref{FIG4} shows the combined topological phase diagram of the two uncoupled wires. The white regions identify the parameter windows in which both wires are topologically nontrivial, whereas the black regions correspond to both wires being trivial. In the orange and cyan regions only the top or bottom wire, respectively, is topological. The interspersed colored regions reflect the different disorder realizations in the two arms. Thus, tuning one wire into a topological regime does not automatically place the other wire in the same regime. Nevertheless, the diagram contains substantial common topological regions, particularly at larger Zeeman fields. These regions provide the natural starting point for implementing the doublon mechanism in a disordered device. The two red cuts allow us to compare the spectra before and after coupling the right ends, first by varying the chemical potential and then by varying the Zeeman field.

\begin{figure}[t]
\begin{center}
\includegraphics[width=0.48\textwidth]{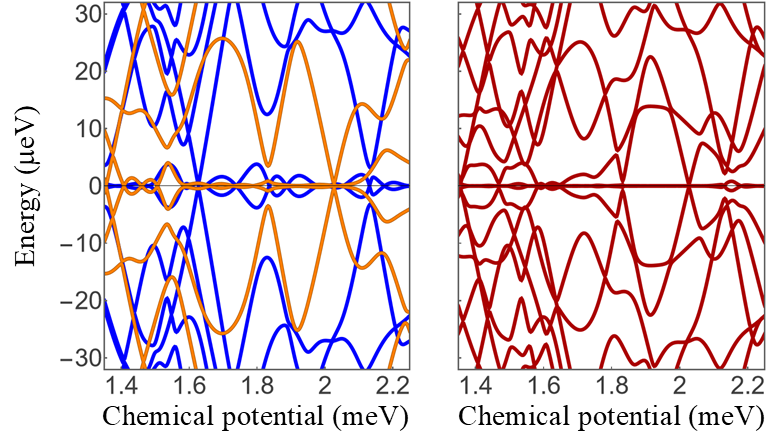}
\end{center}
\vspace{-2mm}
\caption{Energy spectra as functions of the chemical potential along the vertical cut in Fig. \ref{FIG4} corresponding to $\Gamma = 1.04~$meV. {\em Left}: Energy spectra for the uncoupled top (blue lines) and bottom (orange lines) wires. {\em Right}: Energy spectrum of the doublon obtained by coupling the wires through a bridge with $W_x=70~$nm and $W_y=150~$nm. Note the Majorana energy splittings on the order of a few $\muup$eV characterizing the uncoupled wires (left panel) and the significantly reduced energy splitting associated with the ``surviving'' left Majorana modes (M1 and M3) supported by the doublon device (right panel). The V-shaped features are associated with (topologically-trivial) Andreev bound states localized within the top or bottom wires.} 
\label{FIG5}
\vspace{0mm}
\end{figure}

The chemical-potential dependence along the vertical cut, at $\Gamma=1.04~$meV, is shown in Fig.~\ref{FIG5}. The left panel superposes the spectra of the uncoupled top and bottom wires. Despite the presence of low-energy states over broad intervals, their energies oscillate on a scale of a few $\muup$eV. Hence, the near-zero modes of the individual wires still have appreciable finite-length hybridization. Coupling the wires through a bridge with $W_x=70~$nm and $W_y=150~$nm reorganizes this low-energy spectrum, as shown in the right panel. Over much of the common topological region, the lowest mode remains much closer to zero, while the hybridized right-end pair is displaced to finite energy. This is the same separation of energy scales found in the clean device, although disorder introduces additional structure in both the lowest mode and the excited-state spectrum.
The V-shaped features in Fig.~\ref{FIG5} arise from topologically trivial Andreev bound states localized within either wire. Their persistence after the bridge is introduced illustrates that coupling the right ends does not remove disorder-induced states throughout the device. Such states can approach the surviving Majorana pair and reduce its separation from other excitations. Consequently, a small lowest eigenvalue alone is insufficient to characterize a useful operating regime. The phase diagram, the separation from other low-energy states, and the spatial structure of the relevant modes must be considered together.

\begin{figure}[t]
\begin{center}
\includegraphics[width=0.48\textwidth]{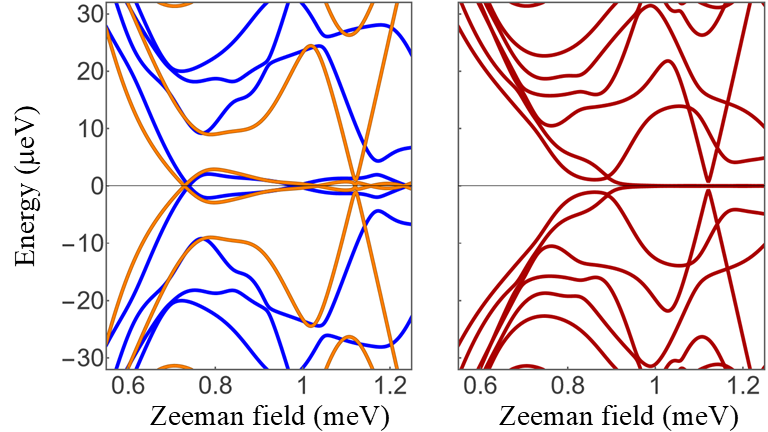}
\end{center}
\vspace{-2mm}
\caption{Energy spectra as functions of the Zeeman field along the horizontal cut in Fig. \ref{FIG4} corresponding to $\mu = 1.9~$meV. {\em Left}: Energy spectra for the uncoupled top (blue lines) and bottom (orange lines) wires. {\em Right}: Energy spectrum of the doublon obtained by coupling the wires through a bridge with $W_x=70~$nm and $W_y=150~$nm. Note that a robust near-zero energy state (corresponding to the Majorana modes M1 and M3) is supported by the doublon (right panel) in the regime where both wires are topologically nontrivial (see Fig. \ref{FIG4}) and $\Gamma \gtrsim 0.9~$meV, i.e., once the energy of the hybridized state M2--M4 is high enough.}
\label{FIG6}
\vspace{0mm}
\end{figure}

A complementary comparison is provided by Fig.~\ref{FIG6}, which follows the horizontal cut at $\mu=1.9~$meV. In the uncoupled wires, the low-energy branches retain finite splitting oscillations as the Zeeman field is increased. In the doublon, a robust near-zero branch emerges for $\Gamma\gtrsim0.9~$meV, within the regime where both wires are topologically nontrivial. The important condition is that the right-end hybridized state has acquired an energy sufficiently large compared with the residual intrawire splittings. At lower fields, where this separation of scales is less favorable, coupling the two arms does not produce the same stable near-zero branch. The figure therefore shows that the benefit of the bridge depends on the operating point as well as on the presence of topological regions in both wires. The additional branches that approach zero at isolated fields remain relevant when selecting a working interval with an adequate excitation gap.

\begin{figure}[t]
\begin{center}
\includegraphics[width=0.48\textwidth]{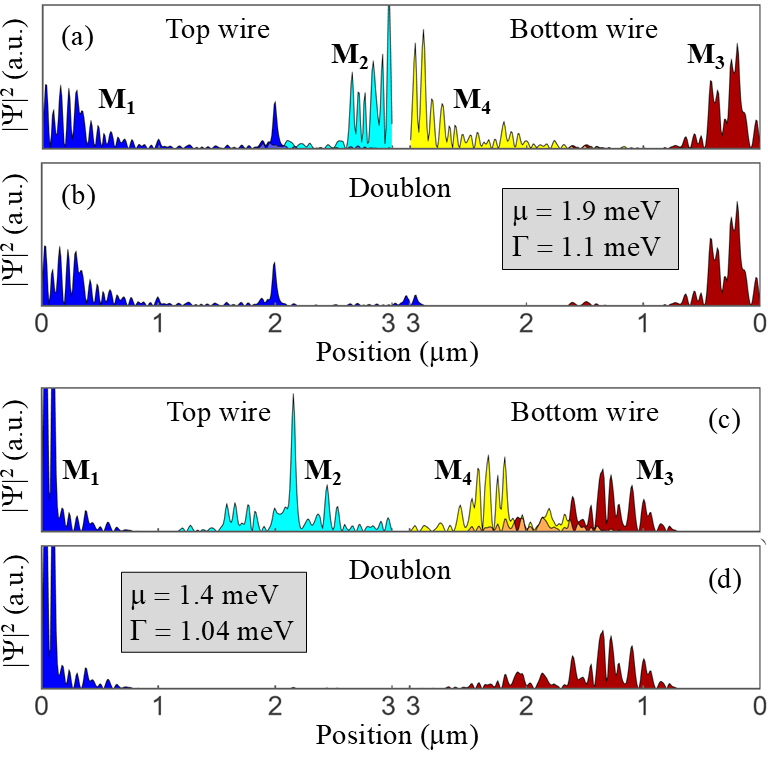}
\end{center}
\vspace{-2mm}
\caption{Spatial profiles of Majorana modes supported by two uncoupled wires, (a) and (c), and by the corresponding doublon devices, (b) and (d), respectively. The spectral weight was integrated across the transverse direction (i.e., the $y$-direction for the wires and the $x$-direction for the bridge). The system parameters are the same as in Figs. \ref{FIG4}-\ref{FIG6} and the specific control parameter values are explicitly given in the figure. Note that the position $x$ along the top wire varies from left ($x=0$) to right ($x=3~\muup$m), while the position along the bottom wire varies from right ($x=3~\muup$m) to left ($x=0$). The small gap between the right ends corresponds to the bridge region of length $W_y=150~$nm. The control parameters in (a) and (b) correspond to both wires being in the topological regime, while for the parameters in (c) and (d) the top wire is in a topological phase and the bottom wire is trivial, with the Majorana modes M3 and M4 hybridizing into a partially separated Andreev bound state. In both cases the coupling through the bridge results in the hybridized mode M2--M4 acquiring finite energy and the ``surviving'' left Majoranas (M1 and M3) becoming well separated.}
\label{FIG7}
\vspace{0mm}
\end{figure}

The spatial profiles in Fig.~\ref{FIG7} clarify the origin of these spectral changes. The horizontal coordinate follows the top wire from its free end toward the bridge and then follows the bottom wire back toward its free end. With this convention, the two joined ends lie next to each other near the center of the plot, while the free ends lie at its two outer edges. The profiles are obtained by integrating the spectral weight over the transverse direction; they therefore show the longitudinal distribution of the Majorana components rather than the full two-dimensional wave functions.

For $\mu=1.9~$meV and $\Gamma=1.1~$meV, both uncoupled wires are topological. Panel (a) shows the four Majorana components, including their disorder-induced secondary peaks and extended tails. The right-end modes M2 and M4 have substantial weight near the bridge and can hybridize efficiently once the arms are joined. Panel (b) shows the two components of the surviving low-energy state in the doublon. M1 and M3 remain concentrated near the free ends, with only small weight along the intervening path. The bridge has thus removed the central pair from the near-zero-energy sector while retaining two well-separated outer components. Their spatial separation explains the reduced splitting seen in the spectra.

Panels (c) and (d) show a different situation, at $\mu=1.4~$meV and $\Gamma=1.04~$meV. Here the top wire is topological, but the bottom wire is trivial. In the latter, M3 and M4 are the partially separated components of an Andreev bound state, rather than Majorana end modes of a topological wire. Coupling through the bridge again hybridizes M2 and M4, leaving a low-energy state with spatially separated M1 and M3 components. However, M3 remains concentrated inside the bottom arm rather than at its free end. This example demonstrates that the spectral reorganization can also occur when one arm supplies a partially separated trivial state. It should not be taken as evidence that both arms are topological or that the resulting mode 
can be used for topological quantum computing, even when its protection is comparable to that of the Majorana pair in panel (b). The practical limitation in this case is generated by M3 being localized away from the free end, which prevents its efficient coupling to a quantum dot.  
For the proposed one-sided tetron, the intended operating regime is the common topological region, with the surviving modes localized near the four free ends.

Taken together, Figs.~\ref{FIG4}--\ref{FIG7} show that the bridge-induced suppression of Majorana splitting survives for the strong disorder realization considered here. Disorder makes the useful regions less uniform and introduces additional low-energy states, but it does not eliminate the basic mechanism. The practical requirement is to find a common topological window in which the bridge hybridizes the joined-end modes strongly and the surviving pair remains separated from other excitations. We next examine how a finite bridge coupling affects the noise-induced parity switching during dot readout.

\subsection{Parity lifetime} \label{SecIIIC}

We now examine the parity lifetime associated with the left Majorana pair using the dot-coupled low-energy Hamiltonian in Eq.~(\ref{HM}). The microscopic calculations presented above establish that a bridge-induced coupling of the M2--M4 Majorana pair on the order of tens of $\muup$eV can be obtained. In the lifetime calculation, we treat $\delta$ as an independent parameter and keep $E_{12}$, $E_{34}$, $t_1$, $t_3$, and the noise strength fixed at the values given in Sec.~\ref{SecII}. This comparison isolates the effect of lifting the right Majorana pair to finite energy and does not assume an additional reduction of the input intrawire splittings (i.e., $E_{12}$ and $E_{34}$) as $\delta$ is increased.

The charge-noise transition rate depends on the matrix element of $H_{noise}$ between the two lowest eigenstates within a fixed total-parity sector. For a given value of the energy splitting $\delta$, the dot energy and the loop flux change the composition of these states and, consequently, their sensitivity to charge fluctuations. At $\delta=0$, both Majorana pairs participate in the low-energy dynamics. A sufficiently large $|\delta|$ separates states involving the right fermion in energy and reduces the mixing that allows dot charge noise to induce transitions between the two lowest states. The relevant suppression in Eq.~(\ref{tauM}) is therefore a suppression of the transition matrix element. Since the calculation uses a fixed effective noise strength, the lifetime enhancement does not require a reduction of the noise amplitude itself.

\begin{figure}[t]
\begin{center}
\includegraphics[width=0.48\textwidth]{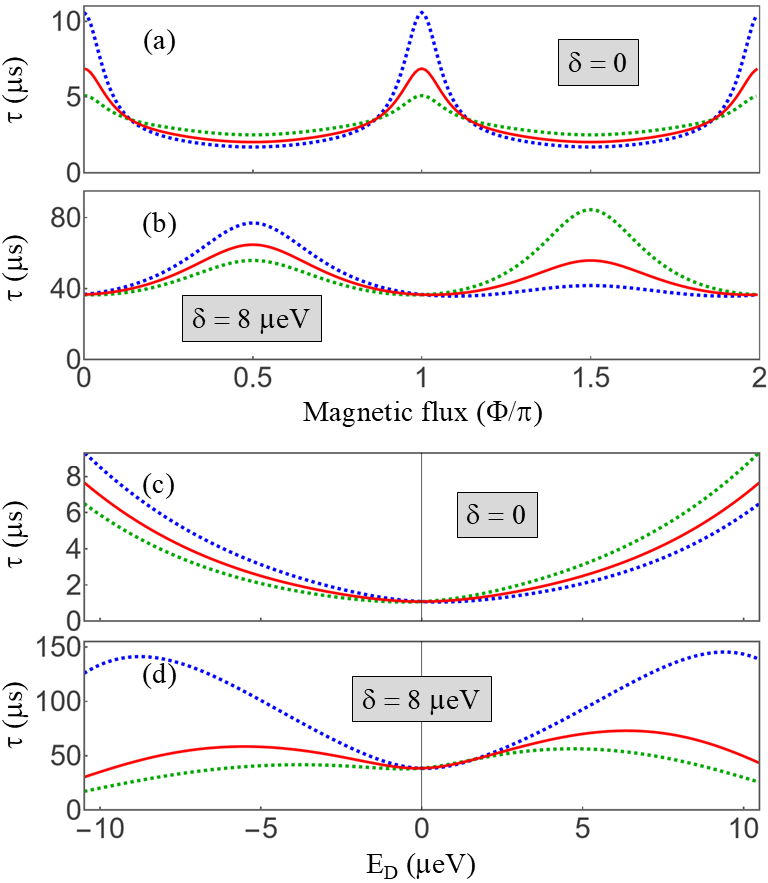}
\end{center}
\vspace{-2mm}
\caption{Dependence of the parity lifetime associated with the left Majorana pair (M1 and M3) on the magnetic flux, (a) and (b), and on the energy $E_D$ of the quantum dot state, (c) and (d). The blue (green) dotted lines correspond to total parity $M=+1$ ($M=-1$), while the red line is the average lifetime (see main text). A decoupled M2--M4 right Majorana pair, which corresponds to $\delta=0$ [panels (a) and (c)], can be realized in a standard tetron device, while a hybridized, finite energy M2--M4 mode with $\delta=8~\muup$eV [(b) and (d)] can be obtained in a doublon device. Note that, upon coupling the M2 and M4 Majorana modes, the parity lifetime associated with the left Majorana pair (M1 and M3) increases by about an order of magnitude. The model parameters used in the calculation are: $E_{12} = 0.15~\muup$eV, $E_{34} = 0.85~\muup$eV, $t_1 = 2~\muup$eV, and $t_3 = 3~\muup$eV. For the top panels the quantum dot energy was set to $E_D = 4~\muup$eV and for the bottom panels the flux is $\Phi=\pi/2$.} 
\label{FIG8}
\vspace{-2mm}
\end{figure}

Figure~\ref{FIG8} compares the dependence of the lifetimes on the relevant control parameters (magnetic flux and dot energy) for $\delta=0$ and $\delta=8~\muup$eV. The blue and green dotted curves give $\tau_+$ and $\tau_-$, while the red curve gives the rate-based average $\tau=2/(\tau_+^{-1}+\tau_-^{-1})$. This average is sensitive to the shorter of the two sector lifetimes; a large lifetime in only one sector does not produce an equally large average. The flux variable in this calculation is expressed in units of $\hbar/e$, as in Eq.~(\ref{HM}), so the biases $\Phi=\pi/2$ and $3\pi/2$ refer to the dimensionless loop phase.

Panels (a) and (b) show the flux dependence at $E_D=4~\muup$eV. Without the bridge coupling, the calculated lifetimes are of order a few microseconds and have maxima near integer multiples of $\pi$. For $\delta=8~\muup$eV, the lifetimes increase to tens of microseconds over the plotted flux interval, with broad maxima near $\pi/2$ and $3\pi/2$. The two sectors respond differently to the flux, and the sector with the longer lifetime changes between these two biases. Nonetheless, both sectors benefit from the finite bridge coupling, and the average lifetime increases by about an order of magnitude. 
The change in the positions of the maxima is also beneficial, since the regimes $\Phi \approx \pi/2$ and $\Phi \approx 3\pi/2$ are the optimal regimes for capacitance measurements \cite{MicrosoftAzureQuantumNature2025}.

Panels (c) and (d) show the dot-energy dependence at $\Phi=\pi/2$. For $\delta=0$, the lifetime is shortest near the dot-charge resonance, $E_D=0$, and increases with detuning. 
Note, however, that increasing the detuning decreases the sensitivity of the capacitance measurement, hence, $E_D$ cannot be arbitrarily large. The sector-resolved curves have different heights and they interchange their relative order as $E_D$ changes sign. 
For $\delta=8~\muup$eV, the lifetime is substantially enhanced even near resonance and develops broad maxima at finite positive and negative detuning. Again, the sector-resolved curves have different heights, but the average lifetime is enhanced relative to $\delta=0$ over the entire plotted interval. Thus, both the flux and the dot detuning provide useful controls, but their optimal regimes are dictated by the sensitivity of the capacitance measurements and, consequently, they cannot be used as efficient knobs for enhancing the parity lifetime. By contrast, a finite, relatively modest hybridization energy induced by the bridge results in a substantial improvement over a wide range of control parameters, including the optimal control parameters regimes. 

\begin{figure}[t]
\begin{center}
\includegraphics[width=0.48\textwidth]{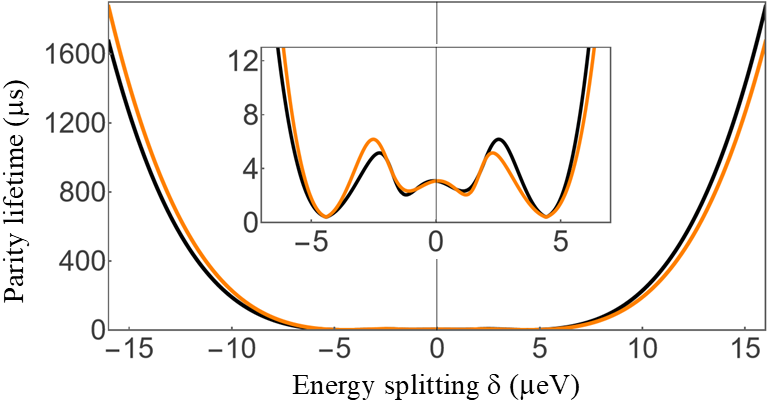}
\end{center}
\vspace{-2mm}
\caption{Parity lifetime associated with the left Majorana pair (M1 and M3) as a function of the energy splitting $\delta$ of the right Majorana pair (M2 and M4). The black (orange) line represents the average lifetime at the optimal flux bias $\Phi=\pi/2$ ($\Phi=3\pi/2$). Note that for small energy splitting values, $|\delta| \lesssim 6~\muup$eV, the parity lifetime has a non-monotonic dependence on $\delta$ and can decrease with respect to the $\delta=0$ value (see inset). By contrast, for energy splitting values exceeding $\sim 10~\muup$eV the parity lifetime increases dramatically by two or three orders of magnitude.} 
\label{FIG9}
\vspace{0mm}
\end{figure}

The full dependence on the bridge coupling is shown in Fig.~\ref{FIG9} for two values of the flux bias, $\Phi=\pi/2$ and $3\pi/2$. The inset resolves the small-$|\delta|$ regime. Here the lifetime is nonmonotonic and can become smaller than its value at $\delta=0$, particularly near $|\delta|\sim 4$--$5~\muup$eV. Therefore, a weak bridge coupling, which generates a low-energy partially-separated Andreev bound state (ps-ABS) with Majorana components M2 and M4, does not guarantee improvement. In this regime, the right-pair energy competes with the dot-induced level separations, changing the ordering and composition of the low-energy states. The charge-noise matrix element can increase before the right pair becomes sufficiently separated in energy to suppress the unwanted transitions. We note that similar physics can be generated by (unwanted) disorder-induced ps-ABSs localized away from the ends of a wire, even when the system is nominally ``topological.'' 

For $|\delta|\gtrsim 10~\muup$eV, the average lifetime rises rapidly for both flux values, reaching values of order a millisecond toward the edges of the plotted range. Relative to the $\delta=0$ lifetime, this corresponds to an enhancement of two or three orders of magnitude. The modest difference between the black and orange curves indicates that either bias can support a large enhancement once the bridge coupling is sufficiently strong. The energy range required for this improvement overlaps the bridge-induced energies obtained in the microscopic calculations. This connection is central to the proposed design: the bridge dimensions can produce a finite-energy right-end state large enough to suppress the charge-noise transition channel, while the free-end pair retains a much smaller residual splitting.

Finally, we note that the calculated lifetime is a property of the dot-coupled system under the assumed noise model, distinct from the coherent precession time of an isolated Majorana qubit, and does not include changes of total parity due to quasiparticle poisoning. The spectral and lifetime calculations presented above show that the proposed one-sided tetron can be a solution for two problems generated by the presence of finite intrawire Majorana splittings. 
On the one hand, a bridge-induced splitting $|\delta|\gtrsim 10~\muup$eV enhances the parity lifetime associated with the ``active'' Majorana pair by up to three orders of magnitude.
On the other hand, the bridge-induced hybridization results in a reduction of the energy splitting of the Majorana modes localized at the free ends, hence in an enhancement of the coherent precession time(s) characterizing the qubit containing these Majoranas. For example, using the values of the intrawire splitting from Fig. \ref{FIG9} with $\delta = 15~\muup$eV results in a hybridization energy for the Majorana modes at the free ends $E_{13} \approx 5.5~$neV. This is smaller that the (average) Majorana splitting characterizing the uncoupled wires, $(E_{12}+E_{34})/2$, by a factor of about 90 and results in a characteristic coherent precession time of about $0.1~\muup$s, as compared to $0.7-0.9~$ns for the uncoupled wires. Most importantly, these benefits can be obtained without improving the quality of the nanowires (e.g., the disorder strength) and without the need to couple to long quantum ``dots'' for performing parity measurements.   


\section{Discussion and conclusions} \label{SecIV}

We have proposed a one-sided tetron -- a qubit device consisting of four parallel topological arms connected at a common end by a trivial superconducting backbone -- as a possible solution for mitigating the adverse effects generated by the intrawire Majorana hybridization, without the need for excessively long nanowire and measurement quantum dots. The qubit is encoded by the four Majorana modes at the free ends. To establish the low-energy mechanisms underlying this geometry, we studied a doublon consisting of two arms connected by a bridge, i.e., half of a one-sided tetron. The doublon geometry clearly isolates the physics associated with the hybridization of the joined-end modes and its consequences for the surviving free-ends pair. 

The microscopic calculations show that the bridge can strongly hybridize the joined-end modes (M2 and M4) while pushing the free-end modes (M1 and M3) close to zero energy. For the clean device, the free-end splitting is reduced by about two orders of magnitude relative to that of an individual arm of the same length. Meanwhile, the hybridized right pair acquires an energy reaching up to $45~\muup$eV. The surviving pair is connected through a path that traverses both arms and the bridge, so a doublon behaves approximately as a folded wire with nearly twice the arm length. This provides the physical basis for achieving the overlap suppression characterizing a longer two-sided device while retaining short individual arms. More importantly, in a one-sided tetron, this mechanism ensures that the distances among the Majorana modes that encode the qubit remain relatively short, so that they can be connected to small quantum dots characterized by widely separated energy levels.  

Strong disorder complicates the parameter dependence, but does not invalidate this mechanism. In general, the two arms have different topological phase maps, and useful operating parameters must be selected within their common topological regions. In those regions, the spectra and spatial profiles demonstrate that coupling the right ends can leave a well-separated free-end pair with substantially reduced splitting. Disorder-induced Andreev bound states localized away from the right end are not affected by the bridge and can reduce the excitation gap at particular operating points, similar to the situations characterizing two-sided devices. 
However, if a partially-separated Andreev bound state emerging in a topologically-trivial arm has one of the Majorana components close to the right end, and the other arm is topological, the low-energy reorganization can occur, generating a pair of nearly zero energy Majoranas. Nonetheless, the Majorana mode localized in the trivial arm is likely to be away from the free end and cannot be efficiently coupled to a quantum dot. This scenario, which is illustrated in in Fig.~\ref{FIG7}(c-d), emphasizes the need to establish the topological character of both wires and the spatial localization of the modes, in addition to simply observing a small energy splitting. In this context, we point out that, in the presence of strong disorder, well separated Majorana modes characterized by extremely low energy splitting but having one (or both) mode(s) localized away from the corresponding end(s) occur generically in long wires.  

The simplified dot-doublon model reveals a second consequence of the bridge-induced hybridization. At fixed intrawire splittings and noise strength, lifting the right Majorana pair to sufficiently high energy suppresses charge-noise-induced transitions between the two lowest states of the doublon-dot system, i.e., enhances the parity lifetime characterizing the dot-left Majorana pair. While the effect is non-monotonic at small values of the bridge coupling, an energy hybridization $\delta=8~\muup$eV gives a lifetime enhancement of about one order of magnitude, while $|\delta|\gtrsim10~\muup$eV allows enhancements of two or three orders of magnitude. Our microscopic calculations show that hybridization energies $|\delta| > 10~\muup$eV are accessible over a wide range of system parameters. The parity lifetime improvement and the reduction of the active (free-end) Majorana splitting are both consequences of effectively doubling the spatial separation of the Majorana modes through the superconducting arms (and bridge), while ensuring that a strong enough bridge-induced hybridization preserves a sufficiently large quasiparticle gap. Since a one-sided tetron consists of two doublons connected trough the common bridge, we expect this mechanism to hold in a four-arm device.   


The bridge geometry must be chosen to generate strong coupling between the Majorana modes localized at the joined ends. In the clean calculations, bridge widths of roughly $60$--$100~$nm consistently support substantial hybridization over most of the control parameter range considered in this study. We find that, all other parameters being fixed, the coupling oscillates with bridge length and its envelope decreases as the bridge becomes longer. 
The bridge parameters used in this study should be viewed as useful starting points, rather than universal design values. Specific wave function characteristics (e.g., the wave number characterizing its oscillatory behavior), electrostatic confinement, proximity coupling, and the relevant control parameter widow may affect the optimal bridge geometry. In addition, electrostatic gates placed near the bridge may provide a means of adjusting it after fabrication. We emphasize that, in general, a useful device must generate a large bridge-induced splitting energy and maintain an adequate gap to all other excitations throughout its intended operating window.

The contrast between the long quasiparticle poisoning-sensitive $Z$-parity lifetime recently reported in InAs--Pb devices and the rather modest improvement of the much shorter $X$-parity lifetime \cite{MicrosoftAzureQuantumNature2025, TetronLifetimes, MicrosoftQuantumPb2026} 
is likely caused by the residual Majorana overlap within each nanowire. In addition, the estimated values of the corresponding Majorana energy splitting ($\sim 0.1-0.3~\muup$eV) generate fast (coherent) isolated qubit dynamics, with a characteristic timescale ($\sim 2-6~$ns) shorter than the typical measurement times, which is not exactly consistent with a topologically protected quantum memory. Even anticipating significant device quality improvements, it is likely that significant suppression of the residual Majorana splitting will require longer nanodevices. For the two-sided family, e.g., the two-sided tetron, this implies widely separated Majorana pairs, which may pose serious problems for parity measurements.
The solution proposed here, based on the one-sided geometry, works efficiently to reduce the Majorana overlap among the MZMs localized near the free ends. Our results show that coupling the joined-end modes to finite energy can reduce the free-end splitting energy by orders of magnitude, while suppressing the quantum dot readout-induced transition mechanism between different parity eigenstates, leading to a dramatic increase of the parity lifetimes. Our calculations support the one-sided tetron as a promising geometry for reducing residual Majorana splitting and improving parity stability without excessively increasing the nominal arm length and the distances among the Majorana zero modes in the qubit.

\section{Acknowledgments}
S.T. acknowledges support from ARO W911NF-22-1-0247. 

\section*{Data Availability Statement}

The numerical data underlying the figures are available from the authors upon reasonable request.

\bibliography{bibliography}

\end{document}